\documentstyle[epsf,twocolumn]{jpsj}

\def\runtitle{Heteroclinic Chaos, Chaotic Itinerancy and Neutral Attractors in
        Symmetrical Replicator Equations with Mutations}
\def\runauthor{Koh {\sc Hashimoto} and Takashi {\sc Ikegami}}

\title{
 Heteroclinic Chaos, Chaotic Itinerancy and Neutral Attractors in
        Symmetrical Replicator Equations with Mutations
}

\author{
Koh {\sc Hashimoto}\footnote{hasimoto@sacral.c.u-tokyo.ac.jp}
and
Takashi {\sc Ikegami}\footnote{ikeg@sacral.c.u-tokyo.ac.jp}
}

\inst{Institute of Physics, College of Arts and Sciences,
      University of Tokyo, 3-8-1 Komaba,Meguro-ku,Tokyo 153,Japan}

\recdate{\today}

\abst{
  A replicator equation with mutation processes is numerically studied.
Without any mutations, two characteristics of the replicator
dynamics are known: an exponential divergence of the
dominance period, and hierarchical orderings of the attractors.
A mutation introduces some new aspects: 
the emergence of structurally stable attractors, and 
chaotic itinerant behavior. 
In addition, it is reported that 
a neutral attractor can exist in the $\mu \rightarrow +0$ region.
}

\kword{chaotic itinerancy, replicator dynamics, Lotka-Volterra equation, heteroclinic cycle, mutation}

\newcommand{\bm}[1]{\mbox{\boldmath$#1$}}

\begin{document}
\sloppy
\maketitle


The replicator equation was initially proposed
by Maynard Smith \cite{msmith},
and was developed thereafter to describe
evolutionary dynamics (see e.g. Hofbauer \cite{sigmund}).
It is equivalent to the Lotka-Volterra equation 
\cite{hofbauer}, and is 
now widely accepted as a basic model equation with applications ranging
from ecosystems 
to other hierarchical network systems.

Recently, unexpected rich behaviors have been found in the equations
\cite{chawanya1,chawanya2}. 
Of these, the emergence of an exponential time scale and 
of a complex attractor hierarchy are worth noting.
The phenomenon of long-time dominance by a unique variable (species) 
and a chaotic transition from one dominant species to 
another are generic features of the equation.
The lifetime of the dominant species is found to diverge exponentially. 
This specific behavior is due to the heteroclinic cycles
embedded in the replicator equation. Depending on one parameter,
 heteroclinic cycles can be hierarchically organized.

On the other hand, the heteroclinic cycle has been considered to be unrealistic
in the light of biological systems.
For example, a population size that decreases to the order of $O(e^{-100})$ is
considered to be unnatural in a real ecosystem.
One remedy for this is to set a lower bound to the population size,
as in the work of Tokita and Yasutomi
\cite{tokita}. 
That is, a species whose population drops below the given threshold must be
removed from the system.
As a result, the system in the end attains a stable distribution of species. 
However, as a consequence, the system loses its rich 
temporal behavior and many degrees of freedom.

In the present letter, we propose another remedy: namely, to recover
structural stability by introducing diffusion terms
(see also
\cite{ikegami,yoshi}).
The diffusion terms can be identified as immigrations and mutations
 in an ecological system. 
We report on the emergence of heteroclinic chaos,
chaotic itinerant phenomena, and neutral attractors
in the replicator equation with mutation.
 


Most studies on relatively large replicator equations
deal with a random interaction matrix
\cite{chawanya1,chawanya2,tokita,ikegami,yoshi}. 
This often makes it difficult to analyze the mechanism controlling
the generic dynamical behavior.
Here we introduce a symmetrical interaction matrix, 
and use a Lotka-Volterra-type equation.
These two types of equations are mutually transformable.
We use a Lotka-Volterra-type equation with seven degrees of freedom as
the simplest model having more than one heteroclinic cycle.
\begin{eqnarray}
\dot{x_i}=x_i \left[ 1 \right. - x_i & - & a(x_{i+1} + x_{i+2} + x_{i+4}) \nonumber \\
                          & - & \left. b (x_{i-1} + x_{i-2} + x_{i-4})\right]
\end{eqnarray}
where $x_i \ge 0  (i=0,1, \cdots , 6)$. 
There are two characteristic parameters $a$ and $b$,
which satisfy the inequalities:
\begin{equation}
a > 1 > b, \;\;\; a-1 > 1-b
\end{equation}

This equation is a natural extension of May's system\cite{may}.
We will discuss a family of replicator-type equations which have several 
properties in common:
1) The $\mbox{\bf{R}}^7_+(=\{\bm{x}|x_i > 0 (i=0,\cdots,6)\})$
plane is kept invariant.
2) There are saddle-type fixed points
$(\bm{e}_i = (0,0, \cdots, 1_i, \cdots, 0))$ at the boundaries 
and one interior repeller-type fixed point, which is    
given by $\bm{q} (= \frac{1}{1+3a+3b}(1,1, \cdots,1))$.
3) A null point gives a trivial fixed point ($\bm{0} = (0,0,\cdots,0)$).

This equation has a symmetrical property in the sense
that every saddle point $\bm{e}_i$ has an equal number of
incoming and outgoing directions of dimensionality one.
For any $i$-th point, there are heteroclinic orbits from $\bm{e}_i$ to
$\bm{e}_{i+j} (j=1,2,4)$.

\begin{figure}[htbp]
\begin{center}
  \epsfile{file=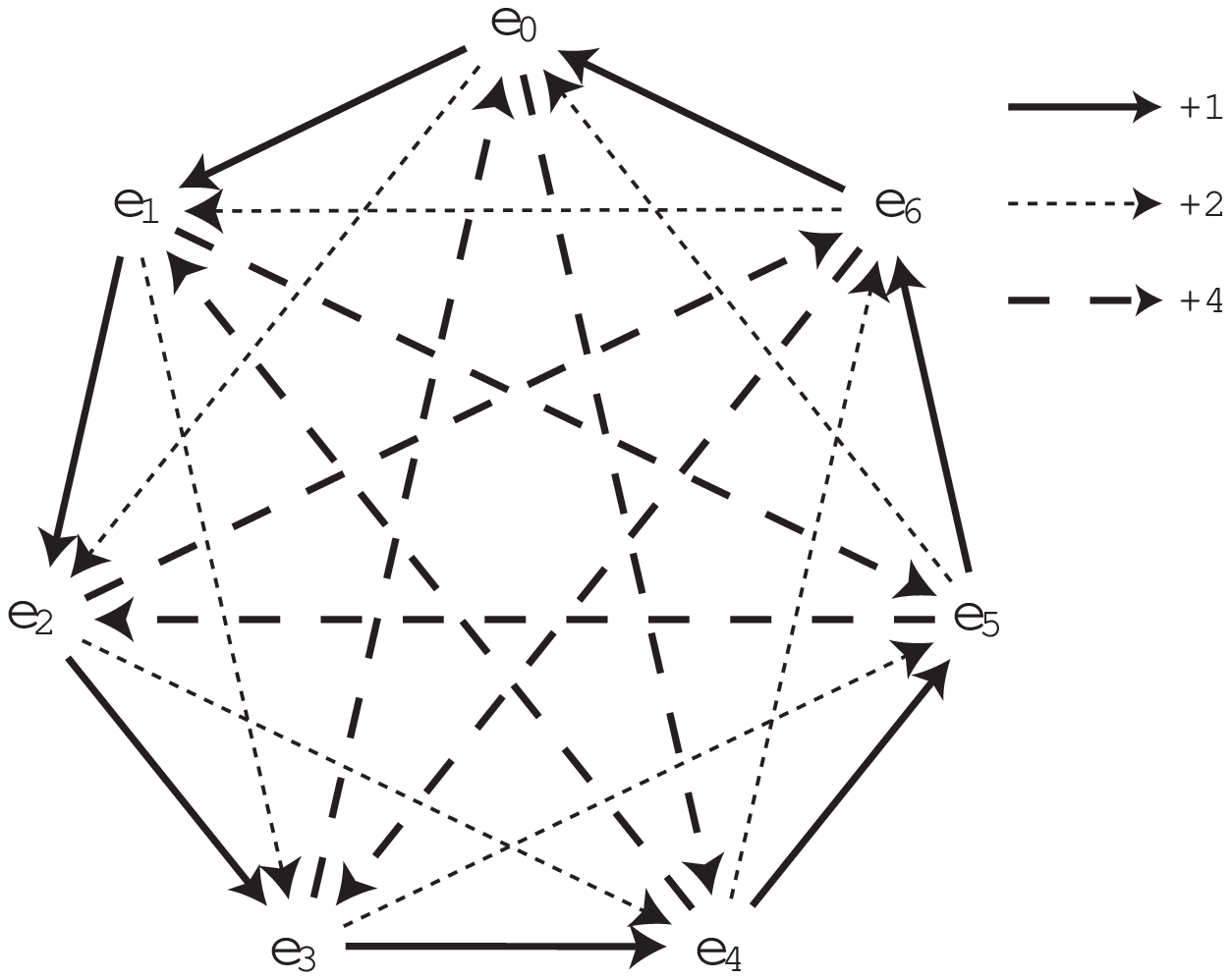,height=4cm}
  \caption{
A diagram of all possible heteroclinic orbits  with seven 
 saddle points symbolized as $\bm{e}_k (k=0,1,\cdots,6)$.  
  Every saddle point has three incoming and 
three outgoing connections to/from other saddle points.
For example,
a saddle point $\bm{e}_1$ has three outgoing connections to $\bm{e}_2$,$\bm{e}_3$ and $\bm{e}_5$
and three incoming connections from $\bm{e}_0$,$\bm{e}_4$ and $\bm{e}_6$.
It is worth noting that $\bm{e}_2$, $\bm{e}_3$ and $\bm{e}_5$
compose another heteroclinic cycle of period 3,
in addition to $\bm{e}_0$, $\bm{e}_4$ and $\bm{e}_6$.
Because of this symmetry, each saddle point is taken to be equivalent.
Thus, there exist three heteroclinic cycles
 which itinerate all the saddle points, but in a different order.
They are drawn in different line styles.
}  \label{fig:cycle7}
\end{center}
\end{figure}

As a result, there are seven saddle points and 21 heteroclinic orbits
in this system.
Each incoming direction is correlated to one of the outgoing directions
of the other saddle point (see Fig.\ref{fig:cycle7}). 
All the initial points, except $x_0 = \cdots x_6$,
will asymptotically converge to a network composed of the 21
heteroclinic orbits.

\begin{figure}[htbp]
\begin{center}
  \epsfile{file=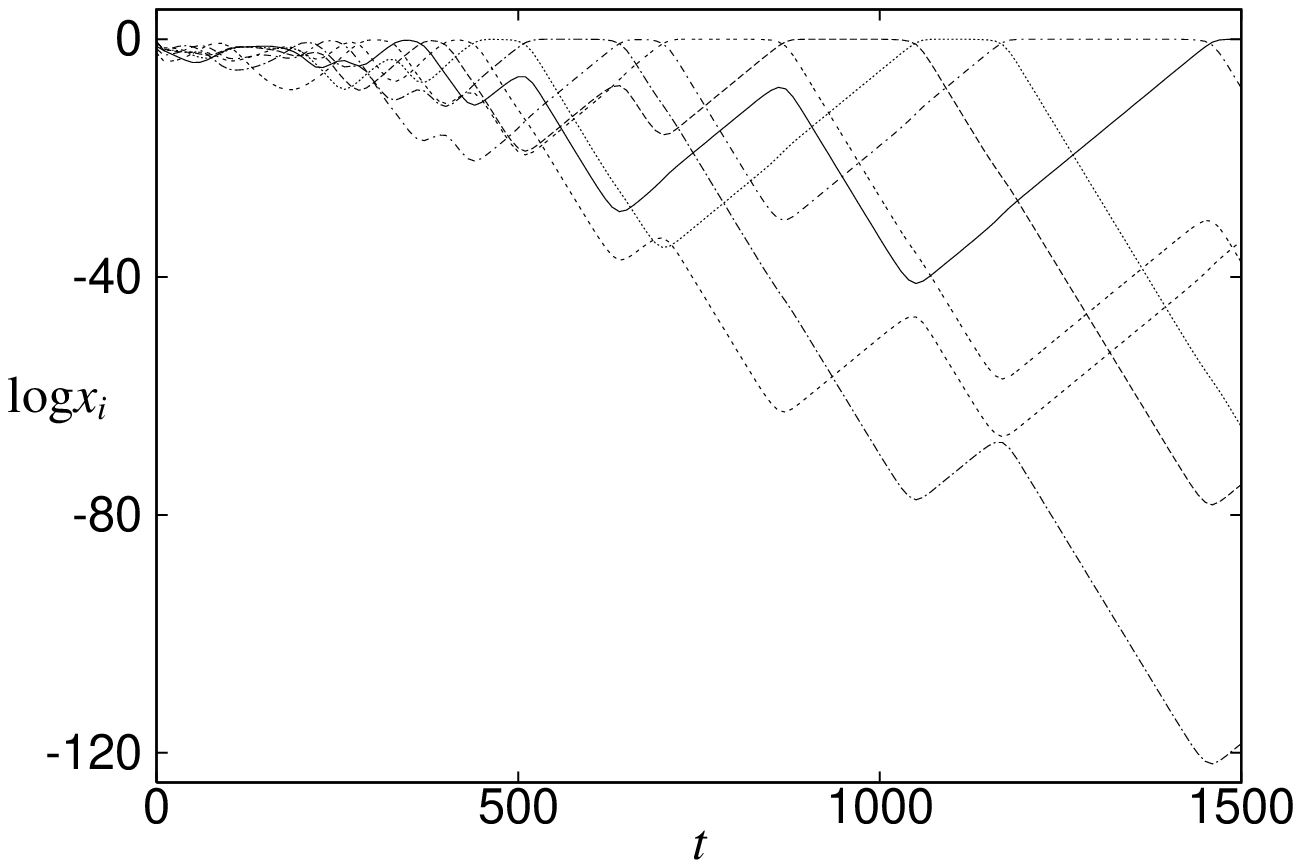,height=2.9cm}
  \epsfile{file=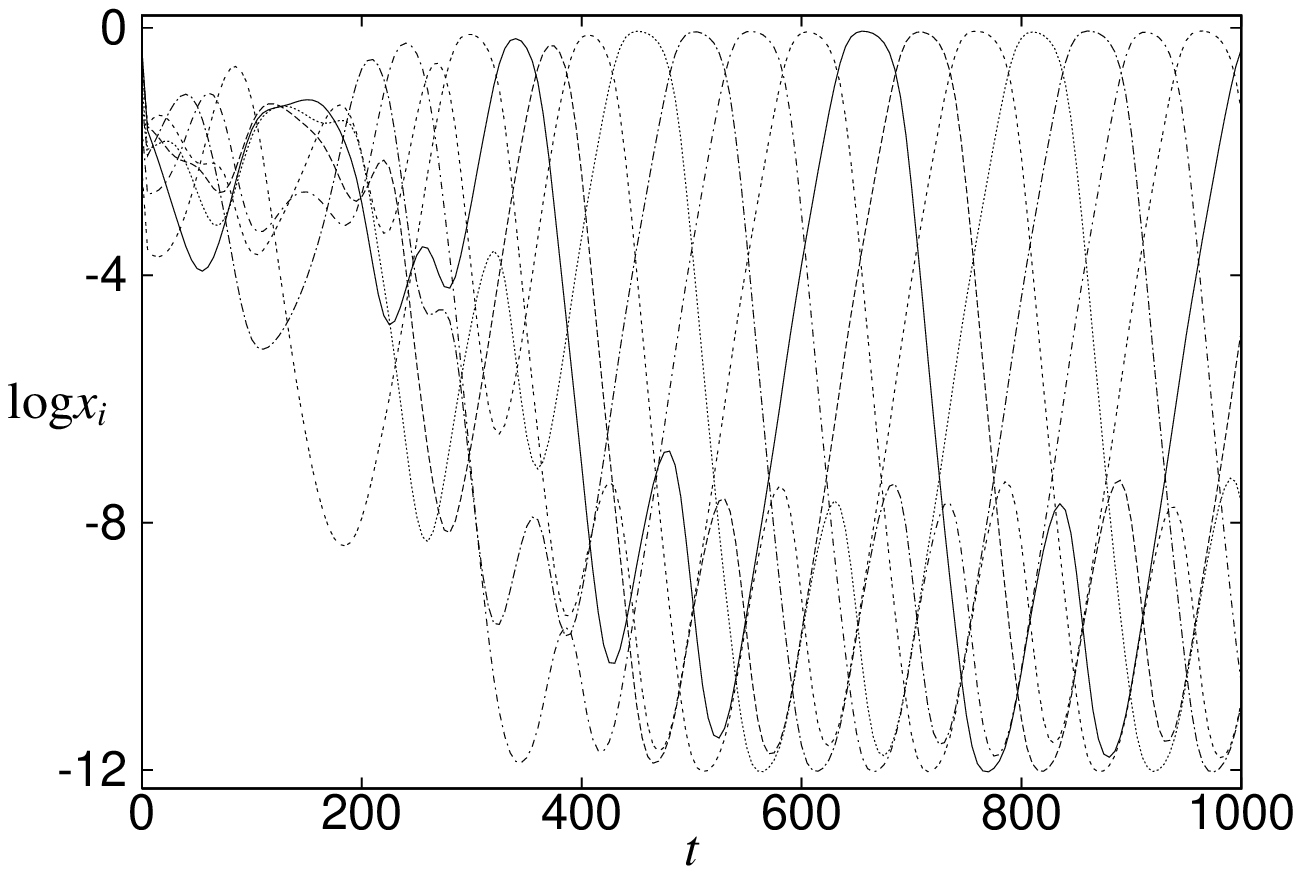,height=2.9cm}
  \caption{
 Time vs. $\log x_i$, simulated with $a=1.2$ and $b=0.9$.
  The dominance periods in the neighborhood of each $\bm{e}_i$
  are gradually extended.
 When $x_i$ is sufficiently close to the maximum value (i.e. unity),
 the growth dynamics of neighboring species are well approximated by 
 the linear curves. That is, they are expressed by $1-b$ and $1-a$ for
           $\frac{d }{d t} \log x_{i+j} (j=1,2,4)$ and
           $\frac{d }{d t} \log x_{i-j} (j=1,2,4)$ ,respectively.
} \label{fig:orbit_nomut}
  \caption{
$t$ - $\log x_{i}$: simulation run with the parameters
    $a=1.2, b=0.9, \mu=10^{-6}$.
           It can be clearly observed that a lower bound exists to each 
population size, which is given approximately by 
 $\frac{\mu}{a-1} \simeq e^{-12.2}$.
} \label{fig:lowbound_orbit}
\end{center}
\end{figure}

\begin{figure}[htbp]
\begin{center}
  \epsfile{file=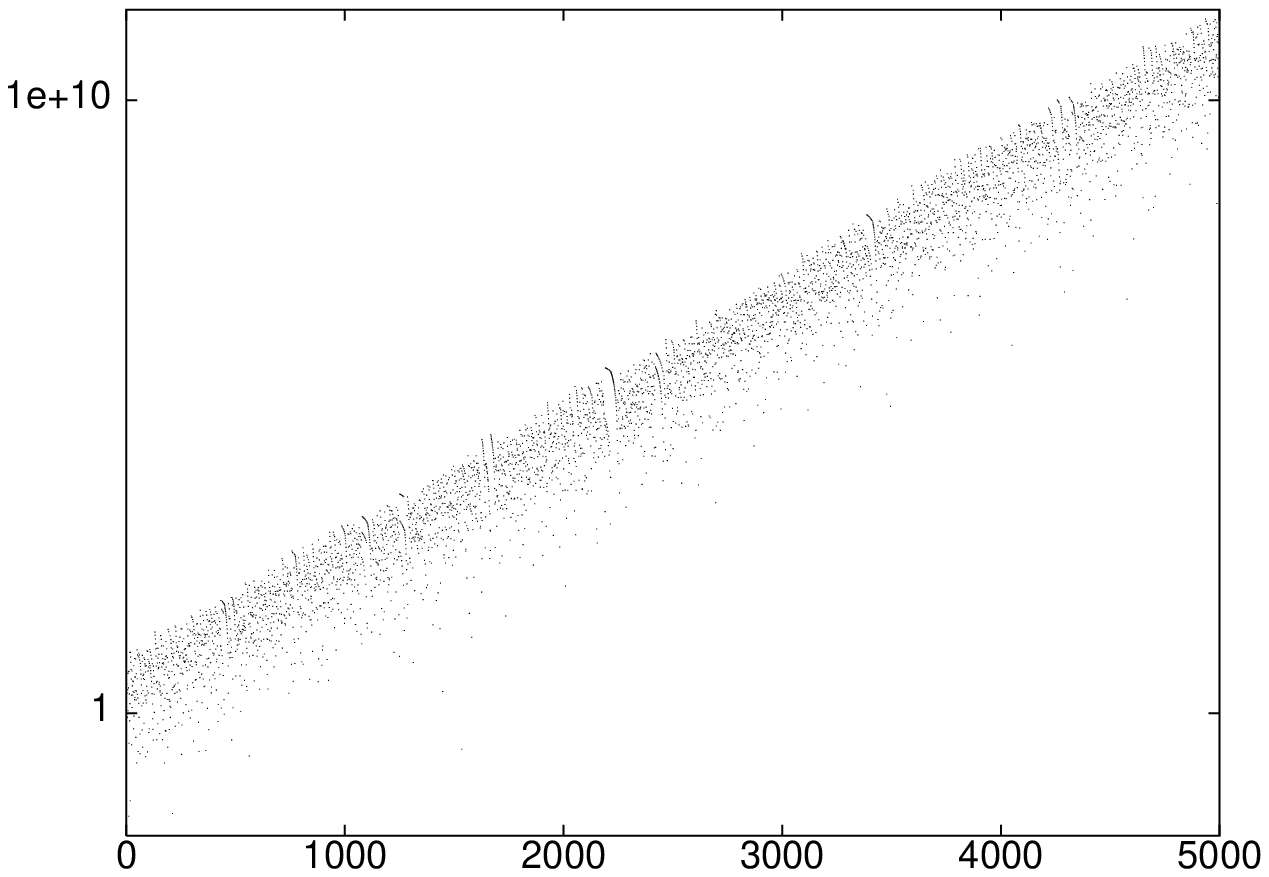,height=4cm,width=7cm}
  \caption{
 The $n$-th recurrence time for visiting a saddle point $\bm{e}_i$
 with parameters $a=1.2$ and $b=0.9$.}
  \label{fig:stay7}
\end{center}
\end{figure}

As was first explicitly pointed out by
Chawanya \cite{chawanya1},
we also numerically find exponential divergence of the 
dominance period in the neighborhood
of saddle points $\bm{e}_i$ even with this symmetric equation
(see Fig.\ref{fig:orbit_nomut},\ref{fig:stay7}).
Two different kinds of transition order
 from $\bm{e}_i$ to $\bm{e}_j$ are obtained.
The transition order from $\bm{e}_i$ to $\bm{e}_j$ becomes either chaotic (a),
or periodic (b) with respect to initial state.
 Almost every case belongs to the chaotic case (a).
In the following, we introduce a diffusion term into the equation. By doing this, any heteroclinic cycle is
removed from the system. Instead, we have the ruins of such cycles and the hoping dynamics among them.


A possible diffusion process for mutation is introduced in the original
 replicator model as follows:
\begin{eqnarray}
\dot{x_i}= & x_i & \left[1 - x_i-a(x_{i+1} + x_{i+2} + x_{i+4}) \nonumber \right.\\
                        &-& \left. b (x_{i-1} + x_{i-2} + x_{i-4}) \right]
                    - 6 \mu x_i + \mu \sum_{j \neq i} x_j.
\end{eqnarray}

Here we assume that there exists mutation from $x_i$ to $x_j$ in the ratio
$\mu$ for all $i$ and $j$.
We call the diffusion process as mutation since it
describes the flow from one population to the others, if we assume
that the variable $x_i$ as the population size of the species $i$ and
all species has the mutually transitionable genotypes.

The mutation process naturally gives
a lower bound $L_{\mu}( \simeq\frac{\mu}{a-1})$
to each population size (Fig.\ref{fig:lowbound_orbit}).
Therefore, every equilibrium point ($\bm{e}_i$) on the peripherals vanishes
simultaneously.
Consequently, heteroclinic cycles no longer exist.
The exponential divergence of the dominance period is also suppressed.

It is also worth noting that introducing mutation rates makes the
system structually stable.  Therefore, the numerical results are
rather insensitive to the numerical method we used. At the same time,
it is more reasonable to use a structually stable model to describe
natural phenomena.  Those two issues are great advantage to study the
replicator equation with diffusion over other models.



The equation is simulated numerically with the 4th order Runge-kutta
method.  Even the system is made structually stable, we solve an
equation of a logarithm of each variable $x_j$. Doing this, we can
study the very lower mutation rate regions.

When the mutation rate $\mu$ is higher than
$\frac{a+b-2}{14(1+3a+3b)}$, the internal equilibrium point $\bm{q}$
becomes stable, which gives a unique fixed point in this system.
Below the critical value, the $\bm{q}$ is destabilized and three limit
cycles appear. Dominant species appear cyclically in each limit cycle
with a fixed but different order($+1, +2$ and $+4$). Each limit cycle
is inversely characterized by this order.
All the initial states in the phase space ($\mbox{\bf{R}}^7_+$)
will be attracted to one of these limit cycles in this parameter region.

\begin{figure}[htbp]
\begin{center}
  \epsfile{file=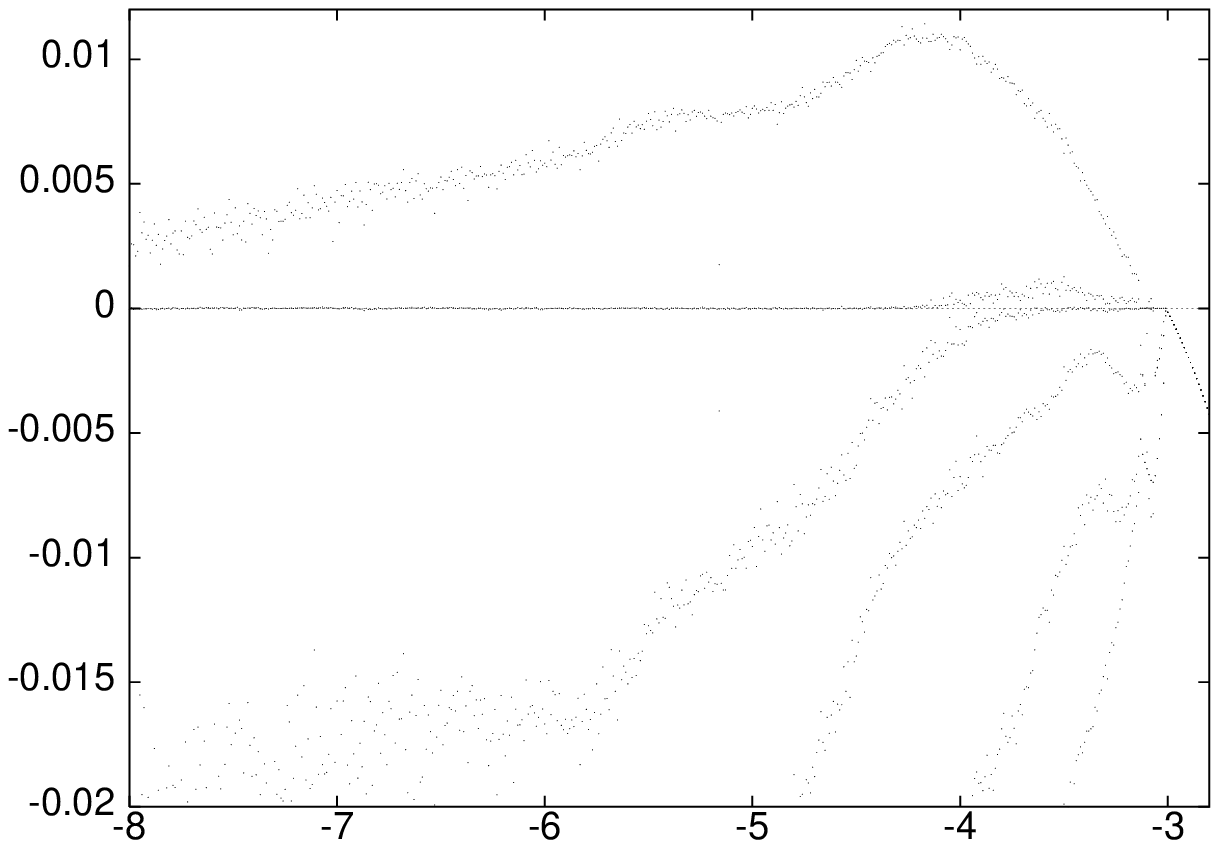,height=4cm,width=7cm}
  \caption{
$\log \mu$ - $\lambda_i$: Lyapunov exponents computed for the parameters
           $a=1.2$ and $b=0.9$ by varying the mutation rates.
}  \label{fig:lyaps}
\end{center}
\end{figure}

By further decreasing the mutation rate, we see sequential period
bifurcation of each limit cycle, each having three quasi-periodic
attractors($T^{2}$) at some point.  Each attractor holds the
``dominant species recurrence order'' which characterizes the original
limit cycles found in the higher mutation-rate regime.  It is
difficult to observe the higher order tori ($T^{n}$), and we instead
observe the breakup of three $T^2$, and the emergence of a strange
attractor. The Lyapunov exponents are computed as a function of the
mutation rate in order to quantify the observed route to chaos in
Fig.\ref{fig:lyaps}.


It is worth noting from Fig.\ref{fig:lyaps} that there are two
qualitatively different chaotic attractors in the higher and lower
mutation regimes: one with two positive Lyapunov exponents, and one
with a single positive exponent.

We can interpret both structures of chaotic attractors as a
combination of three local attractors connected by chaotic dynamics.
These local attractors correspond to the heteroclinic cycles embedded
in the original replicator system without mutations.

The term {\sl chaotic itinerancy} is used to describe a situation
where there is a natural separation of local attractors and chaotic
dynamics which connects those attractors via a higher dimension
subspace\cite{kaneko1}. That is, a state itinerates chaotically among
local attractors.

If the local attractor itself is chaotic, it is difficult to
distinguish whether the state is in or out of local attractors.  This
case is typified by a chaotic attractor with two positive
exponents. On the other hand, if the local attractor is not chaotic,
it is possible to specify the state simply from the time-evolution of
the size of the population. When a state is outside the local
attractors, the population size changes rather chaotically. Indeed, we
observe that the three local strange-attractors degenerate to
quasi-periodic states in the lower mutation regime, where the dynamics
connecting the local attractors remains strange.  Therefore, chaotic
behavior and quasi-periodic behavior appear alternatively.

This picture is well demonstrated by computing the local Lyapunov
exponents.  While the system stays in the local quasi-periodic state,
the three largest exponents are computed as $(0,0,-)$.  While in the
transition states, the exponents are computed as $(+,0,-)$. Further,
we estimate the local Kolmogorov-Sinai(KS) entropy \cite{ksentropy} to
quantify the effective degrees of freedom at each moment.  The local
KS entropy is estimated by the sum of the local Lyapunov exponents
with positive values. The local Lyapunov exponents are computed from
the local Jacobian of the dynamics.

\begin{equation}
(local)KSE = \sum_{n=1}^{q}\lambda_n \; (\lambda_q >0,\lambda_{q+1} \le 0)
\end{equation}

\begin{figure}[htbp]
\begin{center}
  \epsfile{file=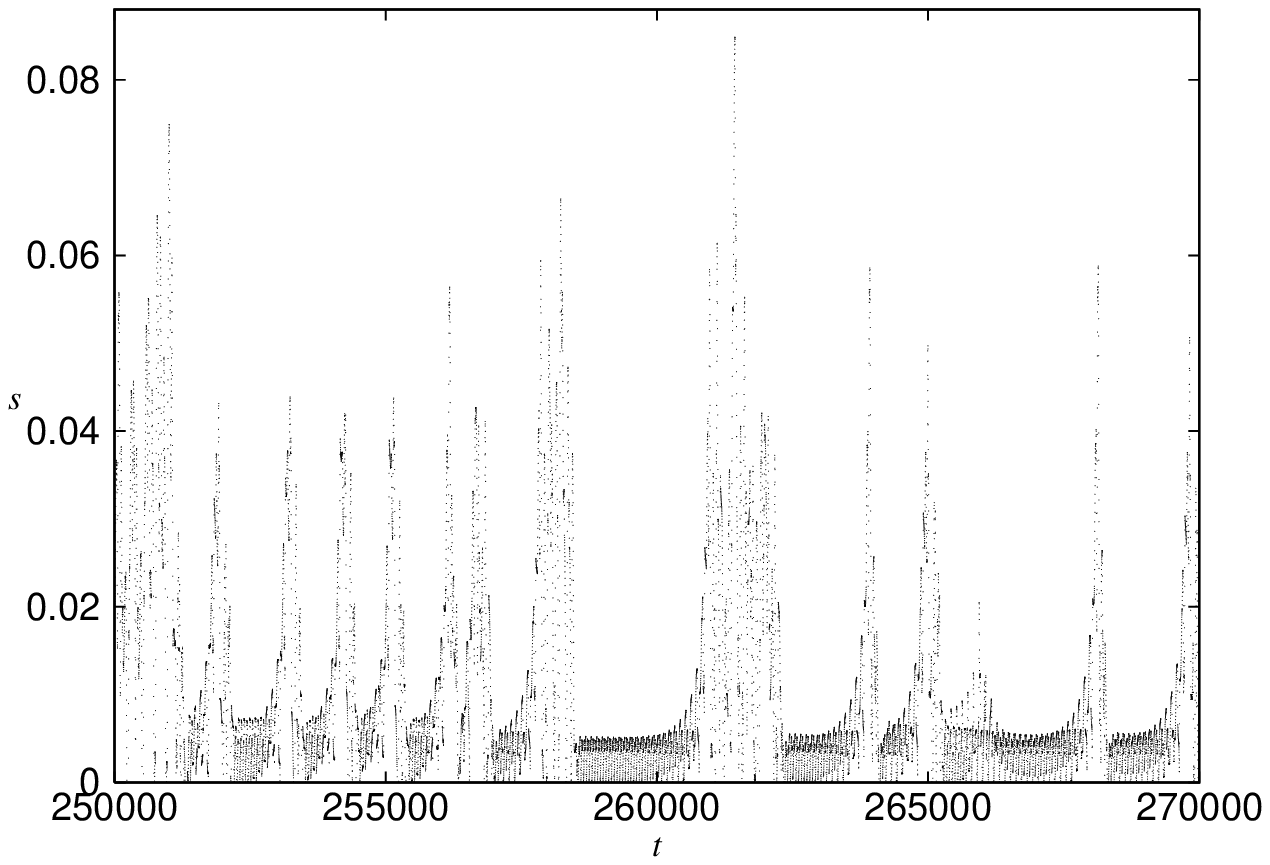,height=4cm,width=7cm}
  \caption{
  $t$ - $local KSE$: Local KS entropy is computed 
           under the condition $a=1.2,b=0.9$ and $\mu=10^{-6}$.
           Where the local KS entropy is near zero,
           the orbit is in the neighborhood of one of the three local attractors.
           On the other hand, where the local KS entropy is large and positive,
           the orbit is in a transition state between local attractors.
}
  \label{fig:local_lyaps_ks}
\end{center}
\end{figure}

The time series of the local KS entropy is plotted in
Fig.\ref{fig:local_lyaps_ks}, where intermittent bursts of the KS
entropy are clearly observed. It is seen from this figure that the
switching between local attractors is associated with a burst of the
local KS entropy. Since the behavior of the local KS entropy is almost
the same as that of the number of the positive local exponents, we
argue that this switching behavior is followed by an increment in the
number of effective degrees of freedom.  From this characterization,
we insist that the switching behavior found in this system should be
named {\sl chaotic itinerancy}.

In the limit of $\mu\rightarrow+0$, the original heteroclinic behavior
is restored. Without any mutation terms, the system can show
exponential growth of a single-species dominance period at $\mu = 0$.
On the other hand, we know that even a small mutation rate can remove
the heteroclinic cycles (i.e. there emerges a lower boundary to a
population size denoted by $L_{\mu}$).  Therefore, we expect that $\mu
= 0$ is a singular point.  When we approach the point from the above
(i.e. $\mu \rightarrow +0$), the largest exponent will approach zero.
This implies that the expected dynamics becomes no more chaotic at
this limit.
Indeed, we see three different periodic behaviors which correspond to
three heteroclinic cycles of the original replicator equation.  Those
periodic behaviors are thus indexed by a recurrent order of the dominant
species(i.e. $+1, +2$ and $+4$). It is worth noting that the motions
with the same recurrence order densly construct a partially
disconnected torus. That is, the limiting behavior is constrained on
the torus.  In the following we analyze the situation.
 
\begin{figure}[htbp]
\begin{center}
   \epsfile{file=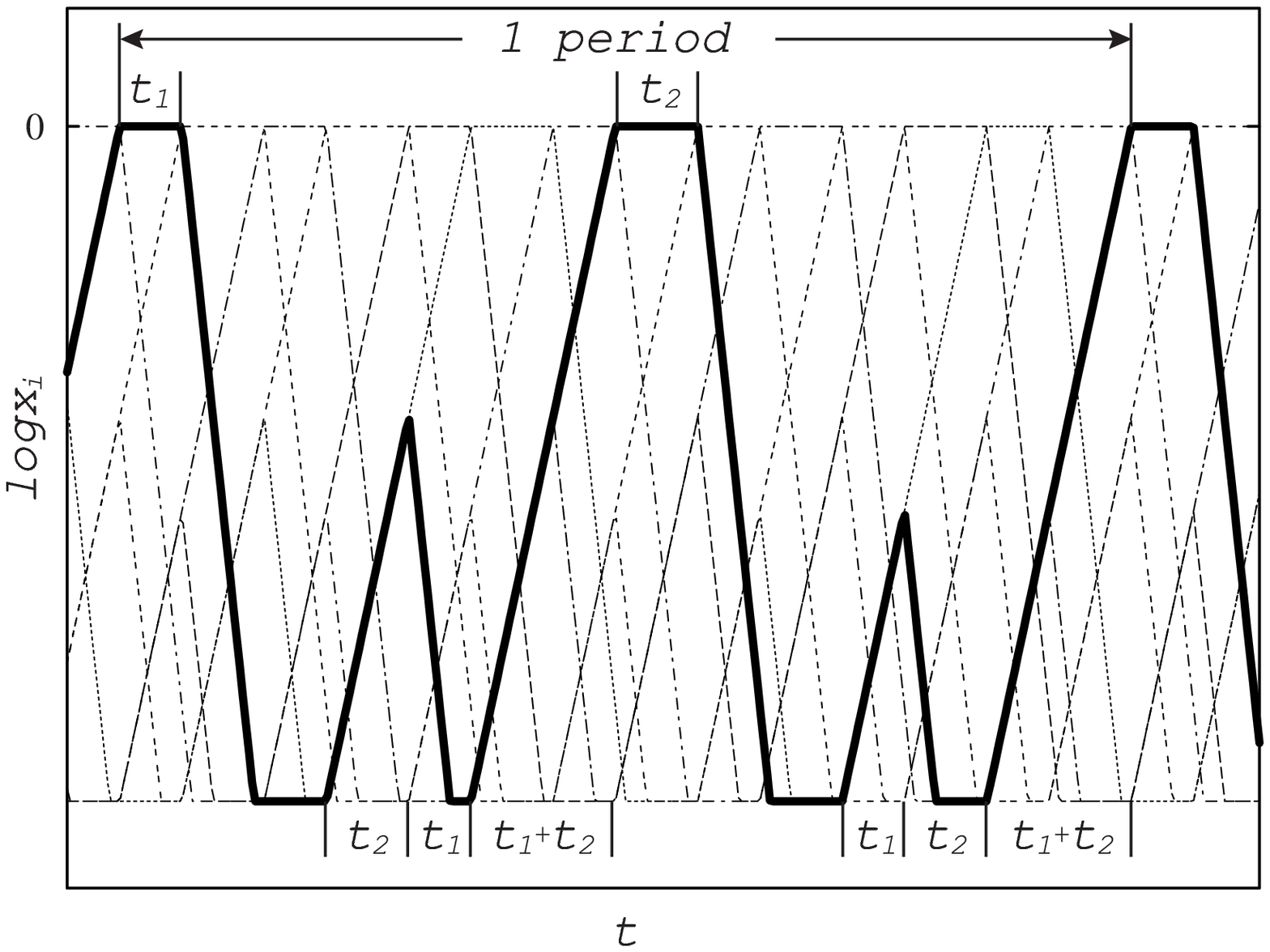,height=3cm,width=6cm}
   \caption{
   With sufficiently small $\mu$,
    the growth factor $\frac{d}{d t}\log x_i$ is approximated by
    $1-a,1-b$ and $0$
    for the incoming, outgoing, and the dominant species, respectively.
}
   \label{fig:y_limit}

\end{center}
\end{figure}

Assuming that the mutation rate is inifinitesimally small, we can
approximate the dynamics by some linearized equations. The limiting
behavior means that the orbit is well charaterized by the transition
between saddle points, and the approximation holds good enough around
saddle points. There, the growth term $\frac{d}{dt} \log x_i$ is well
approximated by $1-a,1-b$ and $0$ for the incoming, outgoing, and the
dominant species, respectively.
It is numerically found that each species becomes dominant twice, and
diminishes (to the lowest order) four times during one period of time
(see Fig.\ref{fig:y_limit}).  This is a necessary feature in order to
keep the periodicity of this linearized equation.  It can be rewritten
more formally in terms of successive dominance periods of given
species, which are denoted by $t_1$ and $t_2$ (see
Fig.\ref{fig:y_limit} for the meaning).
\begin{equation}
(1-b) t_2 + (1-a) t_1 \leq 0,
\end{equation}
\begin{equation}
(1-b) t_1 + (1-a) t_2 \leq 0,
\end{equation}
\begin{equation}
t_1 + t_2 = -\frac{\log L_{\mu}}{(1-b)}.
\end{equation}

The first and second inquality is derived by the fact that each species 
has to be at the lower boundary before it starts to dominate the
population again. The third term is derived in order to hold the
periodicity. That is, 
the sum of $t_1$ and $t_2$ should be conserved by the dynamics, 
but its ratio $\frac{t_2}{t_1}$ can be redundant
under the following condition:
\begin{equation}
1 \leq \frac{t_2}{t_1} \leq \frac{a-1}{1-b} \;\;
(t_1 \leq t_2).
\label{eq_t1t2condition}
\end{equation}

Within this range, absolute values of $t_1$ and $t_2$
can be freely determined.
This degree of freedom gives
a neutral direction of the  periodic state.That is, the periodic state
can exist infinitely many. 

\begin{figure}[htbp]
\begin{center}
  \epsfile{file=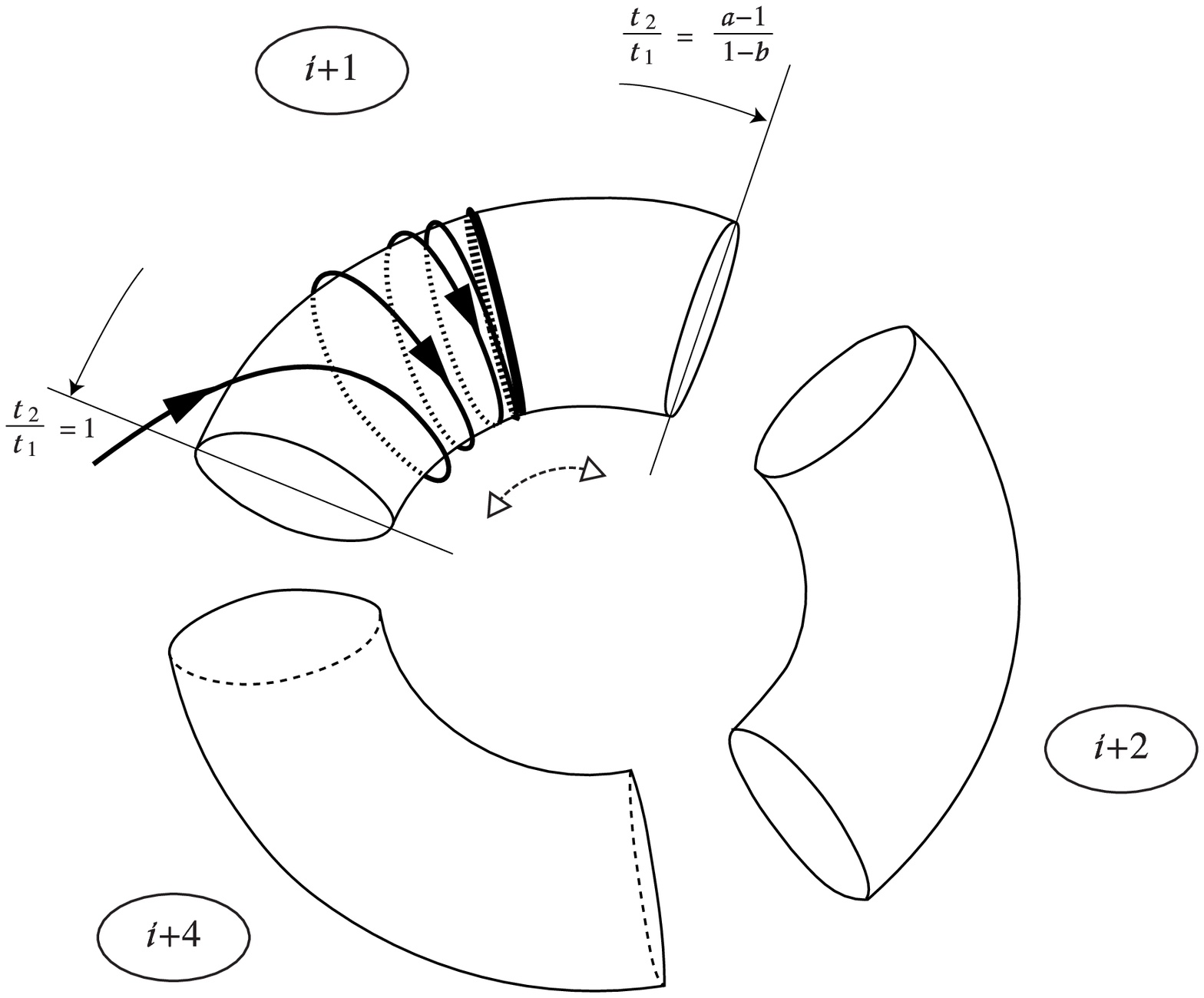,width=4cm}
  \epsfile{file=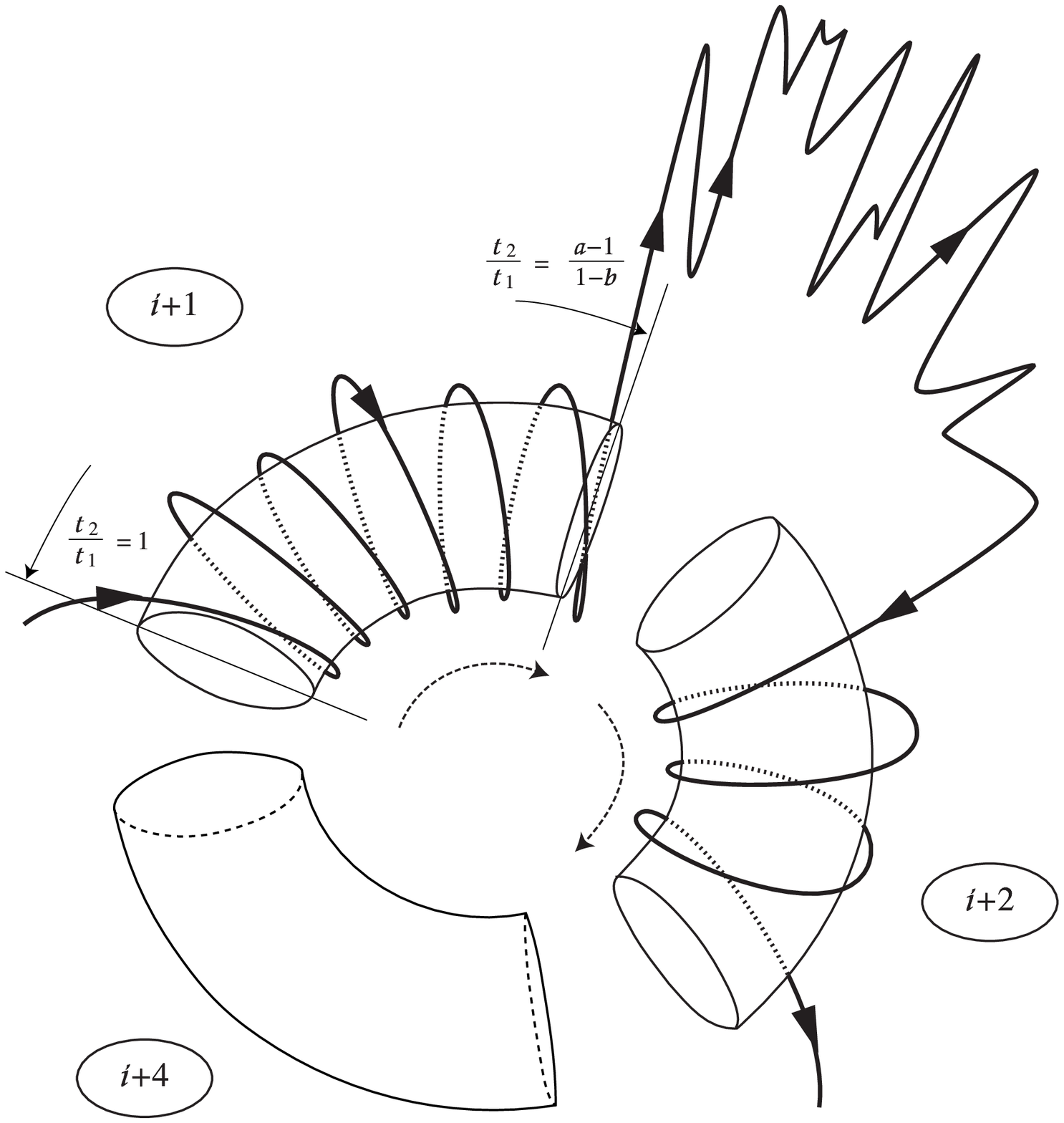,width=4cm}
(a)\hspace{4cm}(b)
  \caption{
  A schematic drawing of the dynamics at $\mu \rightarrow +0$ (a)
and at $\mu = +\epsilon$ (b).
  An orbit asymptotically converges to a cycle on a transverse direction of
  this torus at $\mu \rightarrow +0$.
  But with a small mutation rate, 
  the orbit will no longer stay in one portion of the torus,
  but switches from one to the other.
  The switching behavior appears to be chaotic.

}\label{fig:torus_neutral}
\end{center}
\end{figure}

The above inequalities further set limits on where those periodic states 
can be found. We already know that there are three kinds of periodic
cycle with respect to the recurrence order of the dominat species
($+1,+2$ and $+4$). Each of those three cycles can exist infinitely and 
being found on the unique torus. However, they have to coexist on the
torus and the boundaries between different cycles are given by the 
inequalities. Namely, $(a-1)/(1-b)$ gives the maximal axial length of the
partial torus size, where each cycle can occupy. Therefore, the torus is 
fragmented into three parts, each corresponds to different cycle state (
in Fig.\ref{fig:torus_neutral}(a)). 

 Even with a small mutation rate, we see that the dynamics takes off
 from the torus. The neutral directions discussed above does not exist
 anymore.
 The dynamics with the finite  mutation rate is illuastrated in 
Fig.\ref{fig:torus_neutral}(b).
The  periodic orbit gradually shifts to the edges of the regions,
 and finally switches off to other regions via chaotic motions.
The original periodic behaviors now become quasi-periodic. 

In this letter, we have shown how mutation dynamics changes the behavior of
the original replicator system. 
In particular, a chaotic itinerancy is noted.
We understand phenomenologically that the dynamic behavior 
of this system is well described by the composition of local attractors, and hoping between them, where the local attractors are ruins of heteroclinic cycles.

This system is made symmetric so that each saddle point has an equal number of incoming and outgoing connections to other saddle points.
Understanding a bifurcation diagram and the limiting behavior was possible
due to the symmetry of the system. 

What we have observed in this system is not restricted to the present symmetrical cases. Replicator systems with partially disordered bimatrix 
have been studied,
and the corresponding behavior has been reported (\cite{yoshi}).
A direct extension from this system is to study a series of symmetric equations
with more than three local attractors.
We would then expect that the chaotic itinerant dynamics could be in two ways:
chaotic transition dynamics in time and in space. 
Not only the duration under the influence of one local attractor becomes chaotic, but so does the selection of
the next-switching attractor. 
A detailed description of these systems including more than three local attractors
will be reported elsewhere. 
\section*{Acknowledgement}
This work is partially supported by the COE project ( ``complex systems theory of life'' ) and 
Grant-in aid (No. 11837003) from the Ministry of Education,
Science, Sports and Culture.


\begin{thebibliography}{99}

\bibitem{msmith}
J. M. Smith:
{\it Evolution and the Theory of Games},
(Cambridge University Press, 1982).

\bibitem{sigmund}
J. Hofbauer, K. Sigmund:
{\it The Theory of Evolution and Dynamical Systems}
(Cambridge University Press, 1988).

\bibitem{hofbauer}
J. Hofbauer:
Nonlinear Analysis {\bf 5} (1981) 1003. 

\bibitem{chawanya1}
T. Chawanya:
Progress of Theoretical Physics {\bf 94} (1995) 163. 

\bibitem{chawanya2}
T. Chawanya:
Progress of Theoretical Physics {\bf 95} (1996) 679. 

\bibitem{tokita}
K. Tokita, A. Yasutomi:
Physical Review E {\bf 60} (1999) 682. 

\bibitem{ikegami}
T. Ikegami, E. Yoshikawa:
{\it Towards the Harnessing of Chaos\/}, ed. M. Yamaguchi(Springer-Verlag Tokyo, 1995) p.63.

\bibitem{yoshi}
E. Yoshikawa:
PhD. Thesis, College of Arts and Sciences,University of Tokyo, 1998.

\bibitem{may}
R. M. May, W. J. Leonard:
SIAM J.Appl.Math. {\bf 29} (1975) 243. 

\bibitem{kaneko1}
K. Kaneko:
{\em Physica D\/} {\bf 54} (1991) 5.

\bibitem{ksentropy}
J. P. Eckmann, D. Ruelle:
Rev. Mod. Phys. {\bf 57} (1985) 617. 




\end{thebibliography}
\end{document}